\documentclass[12pt]{article}
\usepackage[top=1in, bottom=1in, left=1in, right=1in]{geometry}
\usepackage{amssymb,amsmath,amsfonts,eurosym,geometry,ulem,graphicx,caption,color,setspace,sectsty,comment,footmisc,caption,natbib,pdflscape,array}
\usepackage{mathptmx}
\usepackage{bm}

\usepackage[title]{appendix}
\usepackage{bbm}
\usepackage{color}  
\usepackage[colorlinks = true,
            linkcolor = black,
            urlcolor  = black,
            citecolor = black,
            anchorcolor = black]{hyperref}
\usepackage{natbib}
\usepackage{tikz}
\usepackage{subcaption}
\usepackage{pgfplots}
\pgfplotsset{width=6cm,compat=1.18}

\usepgfplotslibrary{fillbetween}

\usepackage{booktabs}
\usepackage{tabularx}
\newcolumntype{Y}{>{\raggedright\arraybackslash}X}

\newcolumntype{L}[1]{>{\raggedright\let\newline\\arraybackslash\hspace{0pt}}m{#1}}
\newcolumntype{C}[1]{>{\centering\let\newline\\arraybackslash\hspace{0pt}}m{#1}}
\newcolumntype{R}[1]{>{\raggedleft\let\newline\\arraybackslash\hspace{0pt}}m{#1}}

\usepackage[T1]{fontenc}
\usepackage[utf8]{inputenc}
\usepackage{setspace}
\usepackage{babel}
\usepackage{blindtext}
\usepackage[utf8]{inputenc}
\usepackage{amssymb}
\usepackage{amsthm}
\usepackage{amsmath}
\newtheorem{theorem}{Theorem}
\newtheorem{corollary}{Corollary}

\newtheorem{proposition}{Proposition}

\newtheorem{lemma}{Lemma}
\newtheorem{claim}{Claim}

\usepackage{array}
\usepackage{footmisc}

\title{Revenue Guarantee Equivalence between Dominant-Strategy and Bayesian Mechanisms\thanks{We thank Benjamin Brooks, Songzi Du, Wei He, Jiangtao Li, Qingmin Liu, Vijay Krishna, Stephen Morris, Mallesh Pai, Xianwen Shi, Joel Sobel, Xun Tang, Karen Wu, and audiences at various seminars and conferences for helpful comments. This paper builds on and substantially revises three earlier working papers: \textit{Distributional Robustness: From Pricing to Auctions} (Bachrach and Talgam-Cohen, 2022; extended abstract at EC'22), \textit{Note on Revenue Guarantee for Second-Price Auctions with Random Reserve} (Chen and Yang, 2022), and \textit{Information-Robust Optimal Auctions} (Zhang, 2022). The present paper supersedes those manuscripts. ChatGPT was used to proofread the paper for consistency and clarity.}}
\author{\small
Nir Bachrach\thanks{Technion - Israel Institute of Technology.
Email: nir.bachrach@campus.technion.ac.il}
\hspace{0.5em}
Yi-Chun Chen\thanks{Department of Economics and Risk Management Institute,
National University of Singapore. Email: ecsycc@nus.edu.sg}
\hspace{0.5em}
Inbal Talgam-Cohen\thanks{Technion - Israel Institute of Technology.
Email: inbaltalgam@gmail.com}
\hspace{0.5em}
Xiangqian Yang\thanks{School of Economics and Trade, Hunan University, China.
Email: yangxiangqian@hnu.edu.cn}
\hspace{0.5em}
Wanchang Zhang\thanks{Corresponding author.
School of Management and Economics,
Chinese University of Hong Kong, Shenzhen.
Email: zhangwanchang@cuhk.edu.cn}
}
\date{\small October 20, 2026}

\begin{document}
\maketitle
\thispagestyle{empty}

\begin{abstract}
In auction design with limited information about bidders’ value distributions, we provide an optimality foundation for dominant-strategy mechanisms. We show that simple dominant-strategy incentive compatible (DIC) mechanisms---either a second-price auction with a random reserve or the standard second-price auction without a reserve---achieve a revenue guarantee equal to the worst-case optimal revenue attainable by Bayesian incentive compatible (BIC) mechanisms, thereby establishing a robust form of DIC--BIC equivalence. (JEL C72, D44, D82, D83.)
\end{abstract}
\newpage
\setcounter{page}{1}
\section{Introduction}

A central benchmark in auction theory is Myerson's optimal auction. In the canonical independent private-values model, \cite{myerson1981optimal} characterizes the revenue-maximizing Bayesian incentive compatible (BIC) mechanism. In the symmetric regular case, the optimum takes a simple form: a second-price auction with a deterministic reserve price. Thus, optimal Bayesian revenue can be attained by a dominant-strategy incentive compatible (DIC) mechanism. Dominant-strategy mechanisms are attractive because the seller can predict bidders' bidding behavior without relying on assumptions about their beliefs. The resulting DIC--BIC equivalence in optimal revenue is important because it provides an optimality foundation for these mechanisms.\footnote{A related but distinct notion of DIC--BIC equivalence concerns 
implementation  under a known distribution of independent private types
(\citeauthor{manelli2010bayesian},
\citeyear{manelli2010bayesian};
Gershkov, Goeree, Kushnir, Moldovanu, and Shi,
\citeyear{gershkov2013equivalence}).}

Myerson's benchmark assumes that the seller knows the distribution of bidders' values. In particular, determining the optimal reserve price requires knowledge of this distribution. In many environments, the seller may lack a fully specified distribution or choose to rely only on coarse information, such as aggregate statistics from past auctions or market data. In this paper, we ask whether the DIC--BIC equivalence has a robust counterpart under such limited information and, if so, which dominant-strategy mechanisms attain the robust benchmark.

We begin by studying two baseline informational environments in the symmetric independent private-values model. In the first, the seller knows  the expected highest value, denoted by $\mu$; in the second, she instead knows  the expected second-highest value, denoted by $\eta$. The two statistics capture different aspects of the auction environment: $\mu$ represents the expected surplus from efficient allocation, whereas $\eta$ reflects bidder competition. Under truthful bidding in a sealed-bid second-price auction without a reserve, they correspond to the mean highest bid and the mean transaction price, respectively.\footnote{Bid order statistics and transaction prices
are widely used in empirical auction analysis; see,
e.g., \citet{athey2002identification},
\citet{haile2003inference}, and
\citet{coey2021scalable}. We discuss our
informational constraints in
Section~\ref{sec:info-discussion}.} A value distribution is feasible if it is consistent with the
seller's information.  The revenue guarantee of a dominant-strategy mechanism
is its lowest expected revenue across feasible value distributions.  The \textit{maxmin DIC revenue} is the largest revenue
guarantee attainable by a dominant-strategy mechanism. The \textit{minmax BIC revenue}
is the lowest, across feasible value distributions, of the optimal Bayesian revenue that the seller could obtain if the distribution were known. The latter is an upper bound on the revenue guarantee of any mechanism satisfying voluntary participation, whether
or not it is a dominant-strategy mechanism.\footnote{For the comparison with general mechanisms, we impose
participation security, as in \citet{brooks2021optimal,brooks2024structure}: each bidder has a
message that guarantees a nonnegative payoff regardless of her value or the
messages of the other bidders. Any mechanism with a zero-payoff opt-out
message satisfies this requirement.}  We say that \textbf{robust DIC--BIC equivalence} holds if the maxmin DIC revenue equals the minmax BIC revenue. In that
case, a dominant-strategy mechanism attains the best possible revenue
guarantee among all mechanisms; we call such a mechanism robustly optimal.

In the first baseline informational environment, we show that robust DIC--BIC equivalence holds and a second-price auction
with a random reserve price (SPA-RR) is robustly optimal (Theorem \ref{t1}). In an SPA-RR,
the seller randomly draws a reserve from a distribution and then conducts a
second-price auction with the realized reserve. The mechanism therefore
retains the second-price format while replacing Myerson's deterministic
reserve with a distribution over reserves.

The second-price auction without a reserve (the standard SPA) is a natural starting point because its expected revenue equals the expected second-highest value, which admits a positive lower bound when the expected highest value is fixed (Proposition~\ref{prop:spa-guarantee}). Nevertheless, it leaves revenue unextracted when a high-value bidder faces weak competition.  A reserve price can raise revenue in such cases by replacing the second-highest value as the winner's payment. A deterministic reserve, however, is vulnerable to distributions that place substantial probability immediately below the cutoff. Randomizing the reserve distributes this exposure across cutoffs: probability mass near any one cutoff affects revenue only for a subset of reserve realizations. 

We prove the  result by constructing matching upper and
lower bounds. For the upper bound, we exhibit a feasible equal-revenue
distribution---a distribution under which every posted price in its support
yields the same expected revenue from a single bidder. A key property of this
distribution is that its virtual value \textit{vanishes} throughout the interior of the
support and is positive only at the upper endpoint. Expected revenue therefore
does not depend on how the object is allocated among interior types: many Bayesian mechanisms attain the same optimal revenue under this
distribution. This optimal revenue provides an upper bound on the minmax BIC revenue. For the lower bound, we exhibit a distribution of reserve prices and show
that the associated SPA-RR earns at least this revenue under every feasible
value distribution.  
The lower-bound argument addresses a \textit{nonconvex} optimization problem: the
joint distributions induced by feasible i.i.d.\ marginals form a nonconvex
family of product distributions, so standard linear-programming arguments do
not apply. We derive a revenue representation for a general SPA-RR in which the value distribution enters only through its cumulative distribution function, without requiring a density. We then choose the reserve distribution so that the feasible equal-revenue distribution minimizes the resulting expression pointwise. 

We next consider the second baseline informational environment and show that robust
DIC--BIC equivalence again holds, but now the standard SPA itself is robustly optimal (Theorem \ref{t:second-order-statistic}).   For every realized value profile, the winner in the standard SPA
pays the second-highest value, so its expected revenue is \textit{constant} across feasible value
distributions and equals $\eta$. A feasible
equal-revenue distribution has optimal Bayesian revenue $\eta$, providing the
matching upper bound. This result extends
to lower-ranked order statistics: if the seller knows  the expected $K$th-highest value for some
$K\geq3$, robust DIC--BIC equivalence holds and the standard SPA remains robustly optimal (Corollary
\ref{cor:lower-order}). 

We then examine a joint informational environment in which the seller
knows both the expected highest and second-highest values, $\mu$ and
$\eta$. We show that robust DIC--BIC equivalence continues to hold, with the robustly optimal mechanism depending on the strength of competition relative to
available surplus (Theorem~3(i)--(ii)). When $\eta$ exceeds a threshold
determined by $\mu$, the standard SPA is robustly optimal, and the worst-case
distribution is an equal-revenue distribution truncated at an interior
upper endpoint, with an atom at that endpoint. Both the maxmin DIC and minmax BIC revenues equal $\eta$, so learning the expected highest value in addition to the
expected second-highest value does not improve the optimal revenue
guarantee. When $\eta$ lies below the threshold, an SPA-RR is robustly
optimal, and the worst-case distribution instead combines an atom at zero
with an equal-revenue tail. The atom at zero permits realizations with positive surplus but no
competition---only one bidder has a positive value---in which the
standard SPA earns zero revenue. Table~\ref{tab:main-results-iid} summarizes our main findings for the symmetric independent private-values model.

\begin{table}[htbp]
\centering
\small
\renewcommand{\arraystretch}{0.5}
\setlength{\tabcolsep}{5pt}
\begin{tabular}{@{}lcc@{}}
\toprule
Seller's information
& Robust DIC--BIC equivalence
& Robustly optimal mechanism \\
\midrule
Surplus, $\mu$
& Yes
& SPA-RR \\
\addlinespace
Competition, $\eta$
& Yes
& Standard SPA \\
\addlinespace
Both $\mu$ and $\eta$
& Yes
& Standard SPA or SPA-RR  \\
\bottomrule
\end{tabular}
\caption{Robust DIC--BIC equivalence and optimal mechanisms}
\label{tab:main-results-iid}
\end{table}

Finally, we relax the i.i.d.\ assumption and allow bidders' values to be
arbitrarily correlated. We show that robust DIC--BIC equivalence holds under either baseline informational constraint. When the seller knows the expected
highest value, an SPA-RR is robustly
optimal (Theorem \ref{thm:corr-highest}).  Nature can eliminate competition altogether: in a worst-case
environment, one bidder follows an equal-revenue distribution while all
other bidders have value zero.  When the seller instead knows the expected second-highest value, the
standard SPA remains robustly optimal (Theorem \ref{thm:corr-second}). Its expected revenue is still
constant across feasible value distributions, and an i.i.d.\
equal-revenue distribution  provides the matching Bayesian upper
bound. 

The rest of the paper is organized as follows. The remainder of this section
discusses the related literature. Section~2 presents the model and discusses the informational constraints. Section~3 studies the two baseline informational
environments. Section~4 studies the joint informational environment in which the seller knows both baseline statistics.
Section~5 allows bidders' values to be arbitrarily correlated. Appendix~A
contains omitted proofs.

This paper builds on the benchmark established by \citet{myerson1981optimal}, who shows that a dominant-strategy mechanism attains the optimal Bayesian revenue. Under a known distribution of independent private types, \citet{manelli2010bayesian} establish DIC--BIC implementation equivalence in interim expected allocations and transfers, while Gershkov, Goeree, Kushnir, Moldovanu, and Shi \citeyearpar{gershkov2013equivalence} establish equivalence in interim expected utilities and ex ante expected social surplus in more general social choice environments. Our notion of equivalence instead requires equality between the maxmin DIC revenue and the minmax BIC revenue under distributional uncertainty. With correlated values, however,   DIC--BIC equivalence in optimal revenue fails in general under a known joint distribution. \citet{cremer1985optimal, cremer1988full} illustrate how Bayesian mechanisms can exploit the known correlation structure to extract the full surplus under generic conditions. Our correlated-value results nevertheless establish robust DIC--BIC equivalence under either baseline informational constraint when the correlation structure is unknown.

This paper is related to work on robustness foundations for dominant-strategy mechanisms. Holding fixed the distribution of bidders' values but allowing their belief hierarchies to vary without imposing a common-prior restriction, \citet{chung2007foundations} show, under a regularity condition, that the optimal dominant-strategy auction maximizes worst-case expected revenue over all mechanisms.\footnote{Related foundations
results include \citet{chen2018revisiting}, who study
dominant-strategy mechanisms in general social choice
environments, and \citet{yamashita2022foundations},
who study ex post incentive-compatible mechanisms
with interdependent values.} By contrast, our minmax BIC benchmark considers the optimal Bayesian revenue in the standard common-prior environment induced by each feasible value distribution and then takes the worst case across distributions. 

This paper also contributes to the literature on robust auction design.\footnote{For a broader perspective on
the motivation and appeal of robust mechanism design,
see \citet{brooks2025simplicity}.} One strand of this literature holds the distribution of values fixed and studies robustness to
bidders' information structures and equilibria. Bergemann, Brooks, and Morris \citeyearpar{bergemann2019revenue} compare revenue guarantees across standard
auction formats, while \citet{du2018robust} and
\citet{brooks2021optimal} study robust auction design in common-value
environments. We instead consider uncertainty about the value distribution in private-value environments. Closely related,
\citet{che2022robustly}, \citet{he2022correlation}, and
\citet{zhang2024correlation} study private-value auctions when marginal
means or marginal distributions are known but the correlation structure is
unknown, and identify SPA-RRs as robustly optimal within their respective mechanism classes.\footnote{\citet{che2022robustly} restricts attention to the class they
term competitive mechanisms; see Section~4 of that paper for the 
definition. The finite-market SPA-RR result of
\citet{he2022correlation} concerns standard dominant-strategy mechanisms, which allocate only to a bidder with
a highest report.
\citet{zhang2024correlation} establishes optimality among all
dominant-strategy mechanisms with two bidders and among standard
dominant-strategy mechanisms with three or more bidders, under the
respective regularity conditions.} In our informational environments, by contrast, the auctions we identify attain the minmax BIC upper bound and are therefore robustly optimal without restricting the mechanism class. Thus, both the second-price format and dominant-strategy implementation emerge as conclusions of the analysis. In particular, our results establish exact robust optimality of the standard SPA with finitely many bidders in several informational environments.\footnote{\citet{he2022correlation} establish asymptotic optimality of the standard SPA under known marginal distributions and unknown correlation.} More broadly, we characterize how the seller's information determines whether the standard SPA or an SPA-RR attains the robust benchmark.

Methodologically, our analysis differs from much of the literature on robust auction design. Problems involving uncertainty about correlation or information structures often admit convex formulations over joint distributions or Bayes-correlated outcomes, allowing the use of linear-programming methods. The i.i.d.\ restriction in our setting does not permit the same approach. We instead use a CDF-based representation of SPA-RR revenue to establish tight lower bounds on revenue and characterize the associated worst-case distributions. In the joint informational environment, these distributions exhibit new configurations of atoms and equal-revenue tails.

\section{Model}\label{sec:model}
\subsection{Auction Environment}
A seller sells a single unit of a good, with the supply cost normalized to zero. There are \( n \geq 2 \) bidders. Each bidder \( i \) has a private value \( v_i \) for the good, independently and identically drawn from a distribution \( F \) over the value space \( V = [0, 1] \). This is the canonical symmetric independent private-values (IPV) model. We impose no specific technical assumptions on \( F \); it can be continuous, discrete, or mixed. Let \( \Delta(V) \) denote the set of all Borel probability measures on \( V \). We denote the order statistics as \( v_{(1)} \ge v_{(2)} \ge \cdots \ge v_{(n)} \).

\subsection{Information}

The seller does not know the value distribution $F$, but has limited
information about it. In the first baseline informational environment, the seller knows  the expected
highest value $\mu\in (0,1)$.\footnote{A related informational assumption appears in \citet{zhang2026gda}, who studies bilateral trade when the designer knows the ex-ante gains from trade (GFT), the first-best surplus. Here, $\mathbb{E}[v_{(1)}]$ plays an analogous role in the auction setting.}  The set of feasible value distributions is
\[
\mathcal F_1(\mu)
:=
\left\{
F\in\Delta(V)
\,\middle|\,
\mathbb E_F[v_{(1)}]=\mu
\right\}
=
\left\{
F\in\Delta(V)
\,\middle|\,
\int_0^1 x\,dF^n(x)=\mu
\right\}.
\]
\noindent In the second baseline informational environment, the seller instead knows the expected
second-highest value $\eta\in(0,1)$:
\[
\mathcal F_2(\eta)
:=
\left\{
F\in\Delta(V)
\,\middle|\,
\mathbb E_F[v_{(2)}]=\eta
\right\}.
\]
\noindent We also consider a joint informational environment in which the seller knows both baseline statistics:
\[
\mathcal F_{1, 2}(\mu,\eta)
:=\mathcal F_1(\mu)\cap\mathcal F_2(\eta)=
\left\{
F\in\Delta(V)
\,\middle|\,
\mathbb E_F[v_{(1)}]=\mu,\;
\mathbb E_F[v_{(2)}]=\eta
\right\}.
\]
In all expectations with subscript $F$, the value profile has joint distribution $F^n$.
\subsection{Robust DIC--BIC Equivalence}\label{sec:revenue-benchmarks}
To define our notion of equivalence uniformly across informational environments, let $\mathcal F\subseteq\Delta(V)$ denote the set of value distributions
consistent with the seller's information.

We first define the maxmin DIC revenue. By the revelation
principle, when restricting attention to dominant-strategy implementation,
it is without loss to consider direct mechanisms. A direct mechanism
$M=(q,t)$ consists of an allocation rule
$q:V^n\to[0,1]^n$ and a payment rule $t:V^n\to\mathbb R^n$.
For a reported value profile $v=(v_1, v_2, \cdots, v_n)$, the allocation probabilities are $q(v)=(q_1(v), q_2(v),\cdots, q_n(v))$ and the payments are $t(v)=(t_1(v), t_2(v),\cdots, t_n(v))$, where $q_i(v)\ge 0 $ and $\sum_i q_i(v)\le 1 $ for every $v\in V^n$. We say that a direct mechanism is a dominant-strategy mechanism if
truthful reporting constitutes a dominant-strategy equilibrium  and every bidder
receives a nonnegative ex-post payoff under truthful reporting. Let
$\mathcal D$ denote the set of dominant-strategy mechanisms. For $M=(q,t)\in\mathcal D$ and marginal value distribution $F$, the seller's
expected revenue is $\operatorname{Rev}(M,F)=
\int_{V^n}
\sum_{i=1}^n t_i(v)\,dF^n(v).$ The revenue guarantee of $M$ is its lowest expected revenue over feasible value distributions: $\inf_{F\in\mathcal{F}}\operatorname{Rev}(M,F)$. The maxmin DIC revenue is the largest revenue guarantee attainable by a dominant-strategy mechanism: 
\[
\underline R^{\mathrm{DIC}}(\mathcal F)
:=
\sup_{M\in\mathcal D}
\inf_{F\in\mathcal F}
\operatorname{Rev}(M,F).
\]

We next define the minmax BIC revenue. For this purpose,  fix
a feasible value distribution and suppose that it is common knowledge. A direct mechanism
$M=(q,t)$ is a Bayesian mechanism under $F$ if truthful reporting
constitutes a Bayes--Nash equilibrium and every bidder receives a nonnegative
interim expected payoff under truthful reporting. Let $\mathcal B(F)$ denote
the set of Bayesian mechanisms under $F$. The optimal Bayesian revenue under $F$
is $\operatorname{OPT}^{\mathrm{BIC}}(F)
:=
\sup_{M\in\mathcal B(F)}
\operatorname{Rev}(M,F)$. The minmax BIC revenue is the lowest optimal Bayesian revenue across feasible value distributions:
\[
\overline R^{\mathrm{BIC}}(\mathcal F)
:=
\inf_{F\in\mathcal F}
\sup_{M\in\mathcal B(F)}
\operatorname{Rev}(M,F).
\]

To see why the minmax BIC revenue bounds the revenue guarantees attainable by general mechanisms, consider any mechanism satisfying participation security and any equilibrium under a feasible value distribution \(F\). Participation security ensures that every bidder receives a nonnegative interim expected payoff. By the revelation principle, the equilibrium outcome can therefore be implemented by a Bayesian mechanism under $F$, whose expected revenue cannot exceed $\operatorname{OPT}^{\mathrm{BIC}}(F)$. Taking the worst case over feasible distributions, no mechanism---under any equilibrium selection---can guarantee more than $\overline R^{\mathrm{BIC}}(\mathcal F)$.

We say that \textit{robust DIC--BIC equivalence} holds if
\[
\underline R^{\mathrm{DIC}}(\mathcal F)
=
\overline R^{\mathrm{BIC}}(\mathcal F).
\]
Under this equality, a dominant-strategy mechanism attaining $\underline R^{\mathrm{DIC}}(\mathcal F)$ is robustly optimal: it achieves the best possible revenue guarantee among all mechanisms.
A feasible value distribution attaining $\overline R^{\mathrm{BIC}}(\mathcal F)$ is a worst-case distribution.

For notational convenience, we write
$\underline R^{\mathrm{DIC}}(\mu)$ and
$\overline R^{\mathrm{BIC}}(\mu)$ when
$\mathcal F=\mathcal F_1(\mu)$, and use analogous notation for
$\mathcal F_2(\eta)$ and $\mathcal F_{1, 2}(\mu,\eta)$.
\subsection{Discussion of the Informational Constraints}\label{sec:info-discussion}
The informational constraints---knowledge
of the expected highest value, the expected second-highest
value, or both---provide theoretical \textit{benchmarks} for examining
the optimality foundation for dominant-strategy mechanisms
under limited information. They are not
intended to suggest that
a seller in practice learns only these statistics
and no other information about the value distribution. Rather,
the analysis focuses on mechanisms that rely only on coarse
and portable summaries of the auction environment, asking
whether robust DIC--BIC equivalence continues to hold under
different informational constraints and, if so, how the
nature of the seller's information shapes the robustly optimal mechanisms.  We also consider information about lower-ranked order statistics in Section 3.2 and relax the i.i.d. assumption in Section 5.

\section{Baseline Informational Environments}\label{sec:main}
\subsection{Knowing the Expected Highest Value}
This section studies the first baseline informational environment in which the seller knows the expected highest value, $\mathbb{E}[v_{(1)}]=\mu$. We first exhibit a feasible value distribution that provides the Bayesian
upper bound. We then exhibit a second-price auction with a random reserve
whose revenue guarantee attains that bound.

\subsubsection{Statement of the Result}

For \(a\in(0,1)\), let \(F_a\) denote the equal-revenue distribution
\[
F_a(x)=
\begin{cases}
0, & x<a,\\[3pt]
1-\dfrac{a}{x}, & x\in[a,1),\\[6pt]
1, & x=1.
\end{cases}
\]
The distribution has an atom of size \(a\) at \(1\). It is equal-revenue because every posted price \(p\in[a,1]\) yields the same monopoly revenue $p\cdot\Pr(v\geq p)=a.$ Let \(a\) be the unique
solution to
\begin{equation*}\label{eq:a-feasibility}
\int_0^1 \bigl[1-F_a(x)^n\bigr]\,dx=\mu.
\end{equation*}
The left-hand side is the expected highest value under \(F_a^n\). It is
continuous and strictly increasing in \(a\), with limits \(0\) and \(1\) as
\(a\) converges to \(0\) and \(1\), respectively. Hence, \(a\) is well
defined and \(F_a\in\mathcal F_1(\mu)\).

We next define the candidate mechanism. In a second-price auction with a random
reserve price (SPA-RR),  bidders submit one-dimensional bids,  and the seller draws a reserve price from a distribution independently of bids. The highest bidder
receives the good if her bid weakly exceeds the reserve and pays the maximum of the reserve
and the second-highest bid. Ties among highest bids are broken uniformly. Since the auction is strategy-proof for every realization of the reserve, so is the SPA-RR.

Let \(H_a\) be the reserve distribution given by
\begin{equation*}\label{eq:reserve-cdf}
H_a(x)=
\begin{cases}
0, & x\in[0,a],\\[6pt]
\displaystyle
\lambda_a
\left(\frac{x}{x-a}\right)^{n-1}
\int_a^x
\left(\frac{t-a}{t}\right)^{n-1}\frac{dt}{t},
& x\in(a,1],
\end{cases}
\end{equation*}
where
\begin{equation*}\label{eq:lambda}
\lambda_a
:=
\frac{(1-a)^{n-1}}
{\displaystyle
 \int_a^1
 \left(\frac{t-a}{t}\right)^{n-1}\frac{dt}{t}}.
\end{equation*}
\(H_a\) is a valid CDF. The verification is provided in Appendix \ref{appendixv}. For illustration, Figure~\ref{fig:H_orderstat} plots this CDF for $n=2$ and \(\mu=0.5\). Let \(D_a\) denote the SPA-RR with
reserve distribution \(H_a\).
\begin{figure}[htbp]
\centering
\begin{tikzpicture}
\pgfmathsetmacro{\aa}{0.1108751772}
\pgfmathsetmacro{\delta}{0.005} 
\pgfmathsetmacro{\xcut}{\aa + \delta}

\pgfmathsetmacro{\lamtwo}{(1-\aa)/(-ln(\aa)+\aa-1)}
\begin{axis}[
    axis lines = left,
    xmin=0,
    xmax=1.3,
    ymin=0,
    ymax=1.3,
    xtick={0, \aa, 1,1.3},
    ytick={0,  1,1.3},
    xticklabels = {$0$, $a$, $1$, $x$},
    yticklabels = {$0$,  1, $H_a(x)$},
    legend style={at={(1.1,1)}}
]
\addplot[black, line width=3pt] coordinates {(0, 0) (\aa, 0)};
\addplot[dashed, black!50, forget plot] coordinates {(1,0) (1,1)};
\addplot[dashed, black!50, forget plot] coordinates {(0,1) (1,1)};
\addplot[
    black!70!black,
    thick,
    domain=\aa:\xcut,
    samples=80,
    forget plot
]
{
    \lamtwo * (
        ((x-\aa)/x)/2
        + ((x-\aa)/x)^2/3
        + ((x-\aa)/x)^3/4
    )
};

\addplot[
    black!70!black,
    ultra thick,
    domain=\xcut:1,
    samples=250
]
{
    \lamtwo * (x*ln(x/\aa) + \aa - x)/(x-\aa)
};

\end{axis}
\end{tikzpicture}
\caption{Reserve distribution $H_a(x)$ for $n=2$ and $\mu=0.5$}
\label{fig:H_orderstat}
\end{figure}
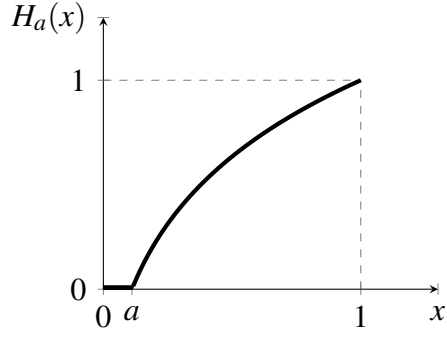
\vspace{2cm}
\begin{theorem}\label{t1}
In the first baseline informational environment, robust DIC--BIC equivalence holds. The mechanism \(D_a\) is robustly optimal, and the value distribution \(F_a\) is worst-case: 
\[
\inf_{F\in\mathcal F_1(\mu)}
\operatorname{Rev}(D_a,F)
=
\underline R^{\mathrm{DIC}}(\mu)
=
\overline R^{\mathrm{BIC}}(\mu)
=
\operatorname{OPT}^{\mathrm{BIC}}(F_a).
\]
\end{theorem}

The theorem follows from two propositions. Define $R:=1-(1-a)^n.$  Proposition~\ref{pwor} shows that
the optimal Bayesian revenue under \(F_a\) is \(R\), providing the Bayesian upper
bound. Proposition~\ref{maxmin} shows that \(D_a\) earns at least \(R\) under
every feasible distribution and exactly \(R\) under \(F_a\).


\subsubsection{The Bayesian Upper Bound}

The following proposition uses \(F_a\) to bound the minmax BIC revenue.

\begin{proposition}\label{pwor}
The optimal Bayesian revenue under \(F_a\) is \(R\):
\[
\operatorname{OPT}^{\mathrm{BIC}}(F_a)=R.
\]
\end{proposition}

\begin{proof}[Proof of Proposition \ref{pwor}]
Under \(F_a\), the virtual value is zero on the continuous part of the
support and equals \(1\) at the upper endpoint. Expected virtual surplus is
therefore bounded above by the probability that at least one bidder has value
\(1\), which is
\[
\Pr\{v_{(1)}=1\}=1-(1-a)^n=R.
\]
This bound is attained by any Bayesian mechanism that allocates the
object with probability one to a bidder with value 1 and has binding lowest-type participation constraints for all bidders. In particular, every second-price auction with a  reserve price  in \([0,1]\), whether deterministic or random,  satisfies these
conditions and therefore earns \(R\).
\end{proof}
\noindent Since \(F_a\) is feasible,  $\overline R^{\mathrm{BIC}}(\mu)\leq R$.

\subsubsection{The Revenue Guarantee of the SPA-RR}

To establish the lower bound, we use the following CDF-based representation of SPA-RR revenue, which allows the worst-case problem to be solved by pointwise minimization of the CDF of the value distribution.\footnote{For expositional convenience, we state
the revenue representation below for atomless
reserve distributions. This restriction suffices
for our lower-bound constructions.} When no confusion arises, we identify an SPA-RR with its reserve distribution $H$
and write $\operatorname{Rev}(H,F)$ for its expected revenue under value distribution $F$.

\begin{lemma}[CDF-based revenue representation]\label{lrp}
For an i.i.d.\ value distribution \(F\) and a reserve distribution \(H\), the expected
revenue of the corresponding SPA-RR is
\begin{equation}\label{eq:single-integral}
\operatorname{Rev}(H,F)
=
\int_0^1
\left[
\underbrace{(1-F(x)^n)\,d(xH(x))}_{\text{marginal allocation efficiency}}
-
\underbrace{H(x)nF(x)^{n-1}(1-F(x))\,dx}_{\text{marginal information rent}}
\right].
\end{equation}
\end{lemma}

\begin{proof}[Proof of Lemma \ref{lrp}]
Let $P$ denote the seller's realized revenue.
Given a value profile $(v_1,\ldots,v_n)$
and a realized reserve price $r$, we have
\[
P
=
\max\{r,v_{(2)}\}\mathbf 1\{v_{(1)}\geq r\}.
\]
For \(x\in[0,1]\),
\[
\Pr(P\geq x)
=
\int_x^1 \Pr(v_{(1)}\geq r)\,dH(r)
+
H(x)\Pr(v_{(2)}\geq x).
\]
Under i.i.d.\ values, $\Pr(v_{(1)}\geq r)=1-F(r)^n$, and $\Pr(v_{(2)}\geq x)=1-F(x)^n-nF(x)^{n-1}(1-F(x)).$ Integrating the tail probability of \(P\) gives
\begin{align*}
\mathbb E[P]
={}&
\int_0^1\int_x^1(1-F(r)^n)\,dH(r)\,dx +
\int_0^1
H(x)
\left[
1-F(x)^n-nF(x)^{n-1}(1-F(x))
\right]dx.
\end{align*}
Changing the order of integration in the first term yields
\[
\int_0^1 x(1-F(x)^n)\,dH(x).
\]
Combining this expression with the term
\(\int_0^1H(x)(1-F(x)^n)\,dx\) yields the revenue representation
in \eqref{eq:single-integral}.
\end{proof}

The representation decomposes revenue into allocation efficiency and information rents. At value level $x$, the reserve must not exceed $x$ for the corresponding surplus to be realized, which occurs with probability $H(x)$. Hence, $xH(x)$ is the surplus generated when the random reserve permits trade, and \(d[xH(x)]\) is the associated marginal surplus. Because this surplus can be realized only when at least one bidder has value above $x$, it is weighted by \((1-F(x)^n)\). Therefore, the first term, \((1-F(x)^n)d[xH(x)]\), is the marginal allocation efficiency. The second term is the marginal information rents. Exactly one bidder has value above $x$ with probability \(nF(x)^{\,n-1}(1-F(x))\). When the reserve is also below $x$, which occurs with probability $H(x)$, this marginal surplus cannot be extracted through either the reserve or bidder competition and accrues to the highest bidder as rents.

\begin{proposition}\label{maxmin}
The revenue guarantee of  the SPA-RR \(D_a\) is $R$, attained by $F_a$:
\[
\inf_{F\in\mathcal F_1(\mu)}
\operatorname{Rev}(D_a,F)=\operatorname{Rev}(D_a,F_a)=
R.
\]
\end{proposition}
\begin{proof}[Proof of Proposition~\ref{maxmin}]
We proceed in four steps.

\medskip
\noindent\textbf{Step 1: Set up the constrained minimization.}

To construct the candidate reserve distribution, we consider an absolutely continuous reserve distribution $H$. By Lemma~\ref{lrp}, the expected revenue of the
associated SPA-RR under an i.i.d.\ value distribution \(F\) is
\[
\operatorname{Rev}(H,F)
=
\int_0^1
\Big[
(1-F(x)^n)\bigl(xH'(x)+H(x)\bigr)
-
H(x)nF(x)^{n-1}(1-F(x))
\Big]\,dx.
\]
A value distribution is feasible if $\int_0^1\bigl[1-F(x)^n\bigr]\,dx=\mu.$ Introducing a multiplier \(\lambda\) on this constraint gives the
Lagrangian
\[
\mathcal L(F;H,\lambda)
=
\operatorname{Rev}(H,F)
-
\lambda
\left(
\int_0^1[1-F(x)^n]\,dx-\mu
\right).
\]
It can be written as
\[
\mathcal L(F;H,\lambda)
=
\int_0^1 \ell_x(F(x);H,\lambda)\,dx+\lambda\mu,
\]
where
\[
\ell_x(y;H,\lambda)
=
(1-y^n)\bigl(xH'(x)+H(x)-\lambda\bigr)
-
H(x)ny^{n-1}(1-y).
\]
Thus, after introducing the multiplier, the value distribution enters the
objective at each price level \(x\) only through the CDF value \(F(x)\).
We construct \(H\) and \(\lambda\) so that \(F_a(x)\) minimizes this
pointwise objective for every \(x\).

\medskip
\noindent\textbf{Step 2: Construct the multiplier.}

Recall that the Bayesian upper bound is
$R=1-(1-a)^n$, where $a$ is determined by
$\int_0^1[1-F_a(x)^n]\,dx=\mu$.
If this upper bound equals the value of Nature's
constrained minimization problem under a suitable
mechanism, the envelope theorem suggests choosing
the multiplier as
\[
\lambda=\frac{dR}{d\mu}.
\]
This envelope calculation motivates the construction; the pointwise
verification below establishes the desired inequality directly and does
not rely on a separate strong-duality result.

Differentiating the feasibility condition and \(R\) with respect to \(a\)
gives
\[
\frac{d\mu}{da}
=
n\int_a^1
\left(\frac{x-a}{x}\right)^{n-1}\frac{dx}{x},
\qquad
\frac{dR}{da}
=
n(1-a)^{n-1}.
\]
Hence,
\begin{equation*}\label{eq:lambda-envelope}
\lambda
=
\frac{dR}{d\mu}
=
\frac{(1-a)^{n-1}}
{\displaystyle
 \int_a^1
 \left(\frac{x-a}{x}\right)^{n-1}\frac{dx}{x}}.
\end{equation*}
Denote it by $\lambda_a$.

\medskip
\noindent\textbf{Step 3: Construct the reserve distribution.}

We next choose \(H\) so that the interior first-order condition for
\(\ell_x(\cdot;H,\lambda_a)\) is satisfied at $F_a(x)=1-a/x$ for $x\in (a,1)$. Differentiating the pointwise objective gives
\[
\frac{\partial\ell_x}{\partial y}
=
ny^{n-2}
\left\{
\bigl[(n-1)H(x)-xH'(x)+\lambda_a\bigr]y
-
(n-1)H(x)
\right\}.
\]
Setting this derivative equal to zero at \(y=1-a/x\) yields
\begin{equation}\label{eq:H-ODE}
xH'(x)
+
\frac{(n-1)a}{x-a}H(x)
=
\lambda_a,
\qquad x\in(a,1).
\end{equation}
We set \(H(x)=0\) for \(x\leq a\). Solving
this differential equation gives
\[
H(x)
=
\lambda_a
\left(\frac{x}{x-a}\right)^{n-1}
\int_a^x
\left(\frac{t-a}{t}\right)^{n-1}\frac{dt}{t},
\qquad x\in(a,1].
\]
Denote it by \(H_a\).

\medskip
\noindent\textbf{Step 4: Verify pointwise minimization.}

We now show that the first-order construction identifies a global minimum. Using \eqref{eq:H-ODE}, for \(x>a\) we obtain
\[
\frac{\partial\ell_x}{\partial y}
=
ny^{n-2}
\frac{(n-1)xH_a(x)}{x-a}
\left[
y-\left(1-\frac{a}{x}\right)
\right].
\]
For $y\in (0,1)$, the derivative is
negative below \(F_a(x)=1-a/x\) and positive above it. Thus
\(y=F_a(x)=1-a/x\) is the unique minimizer on $[0,1]$. For \(x<a\), \(H_a(x)=0\),  so $\ell_x(y;H_a,\lambda_a)
=
-\lambda_a(1-y^n)$, uniquely  minimized at \(y=0=F_a(x)\).  Integrating the pointwise inequalities  gives
\[
\operatorname{Rev}(D_a,F)
\geq
\operatorname{Rev}(D_a,F_a)=R
\qquad
\text{for every }F\in\mathcal F_1(\mu).
\]
The last equality follows from Proposition \ref{pwor}. Since $F_a$ is feasible, the lower bound is attained.
\end{proof}
Since $D_a$ is a dominant-strategy mechanism, $\underline R^{\mathrm{DIC}}(\mu)\geq R$. Propositions~\ref{pwor} and ~\ref{maxmin}, together with the inequality $\underline R^{\mathrm{DIC}}(\mu)\le \overline R^{\mathrm{BIC}}(\mu)$, prove Theorem~\ref{t1}. The strict pointwise inequalities also imply that $F_a$ is the unique distribution minimizing the expected revenue of $D_a$. It follows that $F_a$ is also the unique distribution attaining the minmax BIC revenue in this baseline.

\medskip
\noindent \textit{Performance of the standard SPA.}
For comparison, consider the revenue guarantee of the standard SPA. Since the standard SPA earns \(v_{(2)}\)
at every value profile, its revenue guarantee is $\inf_{F\in\mathcal F_1(\mu)}
\mathbb E_F[v_{(2)}]$. Let \(c_\mu:=(1-\mu)^{1/n}\) and define $\underline\eta(\mu)
:= \mu-n(1-c_\mu)c_\mu^{n-1}.$
\begin{proposition}\label{prop:spa-guarantee}
In the first baseline informational environment, the revenue guarantee of the standard SPA is
\[
\inf_{F\in\mathcal F_1(\mu)}
\mathbb E_F[v_{(2)}]
=
\underline\eta(\mu).
\]
The unique minimizing distribution is the Bernoulli distribution that assigns probability
\(c_\mu\) to zero and \(1-c_\mu\) to one.
\end{proposition}
\begin{proof}[Proof of Proposition \ref{prop:spa-guarantee}]
The proof is provided in Appendix~\ref{app:spa-guarantee}.   
\end{proof}
Proposition~\ref{prop:spa-guarantee} illustrates both the
strength and the limitation of the standard SPA. Under
i.i.d.\ values, fixing a positive expected highest
value imposes a strictly positive lower bound on the
expected second-highest value, so Nature cannot eliminate
bidder competition altogether. Nevertheless, Nature can
weaken competition by concentrating values at the
endpoints of the value space. This leaves scope for
reserve pricing to extract additional revenue
when a high-value bidder faces little competition, as
exploited by the SPA-RR in Theorem~1. For example, when
$n=2$ and $\mu=0.5$, the standard SPA guarantees
approximately $0.0858$, whereas the SPA-RR in
Theorem~1 guarantees approximately $0.2095$.

\subsection{Knowing the Expected Second-Highest Value}
\label{sec:known-second}
We next consider the second baseline informational environment in which the seller knows  the expected
second-highest value, $\mathbb E[v_{(2)}]=\eta$. In contrast to the
first baseline, the robustly optimal mechanism is simply
the standard SPA.

For $\eta\in(0,1)$, define $a_\eta:=1-(1-\eta)^{1/n}$, and let $F_{a_\eta}$ denote the equal-revenue distribution defined in
Section 3.1.1, in which $a$ is replaced by ${a_\eta}$.  Under $F_{a_\eta}$, the expected second-highest value is
\begin{align*}
\mathbb E_{F_{a_\eta}}[v_{(2)}]
&=
\int_0^1
\left[
1-F_{a_\eta}(x)^n
-nF_{a_\eta}(x)^{n-1}
\bigl(1-F_{a_\eta}(x)\bigr)
\right]dx\\
&=
a_\eta
+
\int_{a_\eta}^1
\left[
1-\left(1-\frac{a_\eta}{x}\right)^n
-n\frac{a_\eta}{x}
\left(1-\frac{a_\eta}{x}\right)^{n-1}
\right]dx\\
&=
a_\eta+
\int_{a_\eta}^1
\frac{d}{dx}
\left\{
x\left[
1-\left(1-\frac{a_\eta}{x}\right)^n
\right]
\right\}dx\\
&=
1-(1-a_\eta)^n
=
\eta.
\end{align*}
Therefore, the distribution $F_{a_\eta}$ is feasible.

\begin{theorem}\label{t:second-order-statistic}
In the second baseline informational environment, robust DIC--BIC equivalence holds. The standard SPA is robustly optimal, and the value distribution $F_{a_\eta}$ is worst-case:
\[
\inf_{F\in\mathcal F_{2}(\eta)}
\operatorname{Rev}(\mathrm{SPA},F)
=
\underline R^{\mathrm{DIC}}(\eta)
=
\overline R^{\mathrm{BIC}}(\eta)
=
\operatorname{OPT}^{\mathrm{BIC}}(F_{a_\eta}).
\]
\end{theorem}

\begin{proof}[Proof of Theorem \ref{t:second-order-statistic}]
For every
$F\in\mathcal F_{2}(\eta)$, $\operatorname{Rev}(\mathrm{SPA},F)
=
\mathbb E_F[v_{(2)}]
=
\eta$, which implies
\[
\underline R^{\mathrm{DIC}}(\eta)\geq\eta.
\]

For the matching upper bound, consider the feasible equal-revenue distribution
$F_{a_\eta}$. By Proposition~\ref{pwor}, $\operatorname{OPT}^{\mathrm{BIC}}(F_{a_\eta})
=
1-(1-a_\eta)^n
=
\eta$. Therefore,
\[
\overline R^{\mathrm{BIC}}(\eta)\leq\eta.
\]
Since $\underline R^{\mathrm{DIC}}(\eta)
\leq
\overline R^{\mathrm{BIC}}(\eta)$, all inequalities hold with equality.
\end{proof}
The standard SPA is also uniquely robustly optimal within the class of second-price auctions with possibly random reserves, as established in Proposition \ref{prop}.

\begin{proposition}\label{prop}
Let \(H\) be the reserve distribution of an SPA-RR. If
\(H\neq\delta_0\), where \(\delta_0\) places probability one on a zero
reserve, then
\[
\inf_{F\in\mathcal F_2(\eta)}
\operatorname{Rev}(H,F)
<
\eta.
\]
\end{proposition}
\begin{proof}[Proof of Proposition \ref{prop}]
The proof is provided in Appendix~\ref{app}.    
\end{proof}
The standard SPA earns exactly \(\eta\) under every feasible value distribution. A positive reserve can raise the winner's payment when only one bidder has a high value, but it may also prevent trade when all bidders have lower values. Nature can exploit this trade-off. If the reserve exceeds \(\eta\) with positive probability, a degenerate distribution at \(\eta\) yields revenue below \(\eta\). Otherwise, Nature can choose a two-point distribution
with upper value $1$ and lower value approaching
the upper endpoint of the reserve distribution's
support, while preserving the expected
second-highest value. The potential screening gain then becomes negligible relative to the revenue lost when all bidders have the lower value. Thus, any nontrivial positive reserve strictly lowers the revenue guarantee. This argument does not establish the same conclusion in the first baseline informational environment, where the standard SPA's expected revenue varies across feasible value distributions.

\medskip
\noindent \textit{Lower-ranked order statistics.} The result for the expected second-highest value extends to any lower-ranked order statistic. Corollary \ref{cor:lower-order} formalizes this extension.
\begin{corollary}
\label{cor:lower-order}
Suppose that the seller knows 
\(\mathbb{E}_F[v_{(K)}]=\eta_K\) for some integer \(3\le K\le n\). Then robust
DIC--BIC equivalence holds. The standard SPA is robustly optimal, and both the maxmin DIC and minmax BIC revenues equal
\(\eta_K\).
\end{corollary}

\begin{proof}[Proof of Corollary \ref{cor:lower-order}]
Since \(v_{(2)}\geq v_{(K)}\) for every
realized value profile, the expected revenue of standard SPA satisfies
$$
\mathbb{E}_F[v_{(2)}]\geq \mathbb{E}_F[v_{(K)}]=\eta_K
$$
under every feasible value distribution $F$. For the upper bound, consider the
feasible degenerate distribution under which every bidder's value equals
\(\eta_K\). The total surplus is then \(\eta_K\), so no Bayesian mechanism can obtain expected revenue exceeding
\(\eta_K\). The maxmin DIC and minmax BIC revenues therefore coincide at
\(\eta_K\).
\end{proof}
The expected second-highest value determines the expected revenue of the standard SPA, whereas lower-ranked order statistics provide only lower bounds on this revenue. These bounds are tight in the worst case.

\section{Joint Information about Surplus and Competition}
\label{sec:richer-information}
We now consider the joint informational environment in which
the seller knows both the expected highest and second-highest
values, $\mathbb E_F[v_{(1)}]=\mu$ and
$\mathbb E_F[v_{(2)}]=\eta$.
By Proposition~\ref{prop:spa-guarantee} and the inequality
$v_{(2)}\leq v_{(1)}$, any feasible pair $(\mu,\eta)$
satisfies $\underline\eta(\mu)\leq\eta\leq\mu.$ For fixed $\mu$, either boundary uniquely determines
the value distribution: it is Bernoulli at the lower
bound and degenerate at $\mu$ at the upper bound.
We therefore focus on the interior case $\underline\eta(\mu)<\eta<\mu$.

Recall that in Section~\ref{sec:main}, the parameter
$a$ of the equal-revenue distribution $F_a$
is uniquely determined by $\mu$. Denote this parameter
by $a_\mu$ and define $\bar\eta(\mu)
:=1-(1-a_\mu)^n.$
This threshold is the expected second-highest value under
$F_{a_\mu}$. For fixed $\mu$, the threshold $\bar\eta(\mu)$
divides the feasible values of $\eta$ into two regions:
the high-competition region, where
$\eta>\bar\eta(\mu)$, and the low-competition region,
where $\eta<\bar\eta(\mu)$. We study these regions
in turn and discuss the threshold case
$\eta=\bar\eta(\mu)$ separately.
\subsection{The High-Competition Region}
\label{sec:high-competition}
For $0<a<b<1$, define
\[
F_{a,b}(x)
=
\begin{cases}
0, & x<a,\\[2pt]
1-\dfrac{a}{x}, & a\leq x<b,\\[6pt]
1, & x\geq b.
\end{cases}
\]
The distribution $F_{a,b}$ is a truncated equal-revenue distribution: it is equal-revenue on $[a,b)$, with the remaining probability
mass $a/b$ placed at $b$.  For every feasible pair $(\mu,\eta)$ in the high-competition region,
there is a unique pair $(a,b)$ such that $F_{a,b}\in\mathcal F_{1, 2}(\mu,\eta).$ The existence and uniqueness of these parameters are established in Appendix~\ref{app:low-parameters1}.

\medskip
\noindent\textbf{Theorem 3(i).}
\textit{In the joint informational environment with
$\bar\eta(\mu)<\eta<\mu$, robust DIC--BIC equivalence holds. The standard SPA is robustly optimal,
and the value distribution $F_{a,b}$ is worst-case:
\[
\inf_{F\in\mathcal F_{1,2}(\mu, \eta)}
\operatorname{Rev}(\mathrm{SPA},F)
=\underline R^{\mathrm{DIC}}(\mu,\eta)
=
\overline R^{\mathrm{BIC}}(\mu,\eta)
= \operatorname{OPT}^{\mathrm{BIC}}(F_{a,b}).
\]
}
\begin{proof}[Proof of Theorem 3(i)]
 Under $F_{a,b}$, the virtual value is zero on the
continuous part of the support and positive at the upper endpoint $b$.
The optimal Bayesian revenue is therefore generated by the atom at $b$:
\[
\operatorname{OPT}^{\mathrm{BIC}}(F_{a,b})
=
b\Pr\{v_{(1)}=b\}
=
b\left[
1-\left(1-\frac{a}{b}\right)^n
\right].
\]
Moreover, under $F_{a,b}$,
\begin{align*}
\mathbb E_{F_{a,b}}[v_{(2)}]
&=
a+\int_a^b
\left[
1-\left(1-\frac{a}{x}\right)^n
-n\frac{a}{x}
\left(1-\frac{a}{x}\right)^{n-1}
\right]dx\\
&=
a+\int_a^b
\frac{d}{dx}
\left\{
x\left[
1-\left(1-\frac{a}{x}\right)^n
\right]
\right\}dx\\
&=
b\left[
1-\left(1-\frac{a}{b}\right)^n
\right]
=
\eta,
\end{align*}
where the last equality follows from
$F_{a,b}\in\mathcal F_{1, 2}(\mu,\eta)$.
Thus, $\operatorname{OPT}^{\mathrm{BIC}}(F_{a,b})=\eta.$ This implies
\[
\overline R^{\mathrm{BIC}}(\mu,\eta)
\leq \eta.
\]
On the other hand,  for every
$F\in\mathcal F_{1, 2}(\mu,\eta)$, $\operatorname{Rev}(\mathrm{SPA},F)
=
\mathbb E_F[v_{(2)}]
=
\eta.$
The standard SPA therefore attains the Bayesian upper bound and is robustly optimal.
\end{proof}

In the high-competition region, knowing $\mu$ in addition to $\eta$ changes the
worst-case distribution but not the robustly optimal
mechanism. Whereas $F_{a_\eta}$ in Theorem~2
has support $[a_\eta,1]$, the distribution $F_{a,b}$
has a narrower support $[a,b]$, with $a>a_\eta$
and $b<1$. This compression reduces the
gap between expected surplus and competition,
while preserving the equal-revenue property
under which the standard SPA remains Bayesian optimal.

\subsection{The Low-Competition Region}
\label{sec:low-competition}
For $a,c\in(0,1)$, let $k:=(1-c)a$ and define
\[
F_{a,c}(x)
=
\begin{cases}
c, & 0\leq x<a,\\[2pt]
1-\dfrac{k}{x}, & a\leq x<1,\\[6pt]
1, & x=1.
\end{cases}
\]
The distribution $F_{a,c}$ has an atom of size $c$ at zero, no probability mass on
$(0,a)$, an equal-revenue tail on $[a,1)$, and an atom of size $k$ at one.
For every feasible pair $(\mu,\eta)$ in the low-competition region, there
is a unique pair $(a,c)$ such that $F_{a,c}\in\mathcal F_{1, 2}(\mu,\eta).$ The existence and uniqueness of these parameters are established in
Appendix~\ref{app:low-parameters2}.

We next define the candidate mechanism. Let $H_{a,c}$ be the reserve distribution given by
\[
H_{a,c}(x)
=
\begin{cases}
0, & 0\leq x\leq a,\\[6pt]
\displaystyle
\lambda_{a,c}
\left(\frac{x}{x-k}\right)^{n-1}
\int_a^x
\left(\frac{t-k}{t}\right)^{n-1}
\frac{t-a}{t-k}\frac{dt}{t},
& a<x\leq 1,
\end{cases}
\]
where
\[
\lambda_{a,c}
:=
\frac{(1-k)^{n-1}}
{\displaystyle
\int_a^1
\left(\frac{t-k}{t}\right)^{n-1}
\frac{t-a}{t-k}\frac{dt}{t}}.
\]
The validity of $H_{a,c}$ is established in
Appendix~\ref{appv2}. Let $D_{a,c}$ denote the
SPA-RR with reserve distribution $H_{a,c}$.

\medskip
\noindent\textbf{Theorem 3(ii).}
\textit{In the joint informational environment with
$\underline \eta(\mu)<\eta<\bar\eta(\mu)$,  robust DIC--BIC
equivalence holds. The mechanism $D_{a,c}$ is robustly optimal,
and the value distribution $F_{a,c}$ is worst-case:
\[
\inf_{F\in\mathcal F_{1,2}(\mu,\eta)}
\operatorname{Rev}(D_{a,c},F)
=
\underline R^{\mathrm{DIC}}(\mu,\eta)
=
\overline R^{\mathrm{BIC}}(\mu,\eta)
=\operatorname{OPT}^{\mathrm{BIC}}(F_{a,c}).
\]
}
\begin{proof}[Proof of Theorem 3(ii)]
The proof is provided in Appendix~\ref{app:thm3}.
\end{proof}

Compared with the worst-case distribution in
Theorem~1, $F_{a,c}$ introduces an atom at zero
and raises the lower endpoint of the positive
tail from $a_\mu$ to $a>a_\mu$.
The atom at zero allows realizations in which
only one bidder has a positive value, thereby
weakening competition, while the higher positive
values preserve expected surplus.

This structure also explains why the standard SPA is
no longer robustly optimal. Under $F_{a,c}$, the virtual value is again zero on the continuous part of
the positive tail and positive only at the upper
endpoint 1. Allocating to a bidder with value zero
generates no revenue from that type but
increases the information rents of higher types. Moreover, raising any reserve below $a$ to $a$
preserves allocation to positive-valued bidders
while increasing revenue.  Consequently, a second-price auction with any reserve in  $[a,1]$
attains the optimal Bayesian revenue
\[
\operatorname{OPT}^{\mathrm{BIC}}(F_{a,c})
=
1-(1-k)^n:=R_{a,c}.
\]
In particular, the SPA with reserve $a$ earns
$R_{a,c}$, whereas the standard SPA earns $\eta$.
The two auctions differ in revenue only when exactly
one bidder has a positive value. In that event,
the standard SPA charges zero, whereas the auction with reserve
$a$ charges $a$. Since each bidder has a
positive value with probability $1-c$, the revenue
gap is
\[
R_{a,c}-\eta
=
an(1-c)c^{n-1}
=
nk c^{n-1}
>0.
\]

For the matching lower bound, the reserve distribution
$H_{a,c}$ extends the construction from the first
baseline informational environment. It is chosen so
that $F_{a,c}$ uniquely minimizes the expected revenue
of $D_{a,c}$ over all feasible value distributions, as established by pointwise minimization. Consequently, $F_{a,c}$ is also the unique distribution attaining the minmax BIC revenue.

At the threshold case $\eta=\bar\eta(\mu)$, we have $c=0$,
so $F_{a,c}$ reduces to the equal-revenue distribution
$F_a$, and the reserve distribution $H_{a,c}$
reduces to $H_a$. Both the standard SPA and the SPA-RR from
the first baseline informational environment are then robustly optimal.

\medskip
\noindent \textit{Relation to fixed-distribution DIC--BIC equivalence.}
The fixed-distribution equivalence results of \citet{manelli2010bayesian} and Gershkov, Goeree, Kushnir, Moldovanu, and Shi \citeyearpar{gershkov2013equivalence} imply that the minmax DIC revenue equals the minmax BIC revenue under independent private values. Our results 
establish the remaining equality: the maxmin DIC revenue
equals the minmax DIC revenue. This equality does not follow
from a standard minimax theorem, since the set of feasible
i.i.d.\ joint distributions is nonconvex. Instead, our proofs
establish it constructively by identifying, in each
informational environment, a dominant-strategy mechanism
and a feasible value distribution that form a saddle point.
\section{Arbitrary Correlation}\label{sec:correlation}
 In this section,  we  drop
the i.i.d. restriction and allow bidders' values to follow an arbitrary joint
distribution.  Let \( \Delta(V^n) \) denote the set of all Borel probability measures on \( V^n \), and let $G\in\Delta(V^n)$ denote a joint distribution of the value
profile $v=(v_1,\ldots,v_n)$. When the seller knows  the expected highest
value, the set of feasible joint distributions is
\[
\mathcal G_1(\mu)
:=
\left\{
G\in\Delta(V^n)
\;\middle|\;
\mathbb E_G[v_{(1)}]=\mu
\right\},
\]
and when the seller instead knows the expected second-highest value, the
corresponding set is
\[
\mathcal G_2(\eta)
:=
\left\{
G\in\Delta(V^n)
\;\middle|\;
\mathbb E_G[v_{(2)}]=\eta
\right\}.
\]

For a joint distribution $G$, let $\mathcal B(G)$ denote the set of Bayesian mechanisms under $G$, and let $\operatorname{Rev}(M,G)$ denote the
expected revenue of mechanism $M$. The maxmin DIC and minmax BIC revenues are defined as in Section~2, replacing the set of feasible i.i.d.\
marginal distributions by $\mathcal G_1(\mu)$ or $\mathcal G_2(\eta)$.
Thus, for example,
\[
\underline R_{\mathrm{corr}}^{\mathrm{DIC}}(\mu)
:=
\sup_{M\in\mathcal D}
\inf_{G\in\mathcal G_1(\mu)}
\operatorname{Rev}(M,G),
\qquad
\overline R_{\mathrm{corr}}^{\mathrm{BIC}}(\mu)
:=
\inf_{G\in\mathcal G_1(\mu)}
\sup_{M\in\mathcal B(G)}
\operatorname{Rev}(M,G),
\]
with analogous notation for $\mathcal G_2(\eta)$. We study the two baseline informational constraints in turn.

\subsection{Knowing the Expected Highest Value}

Suppose the seller knows 
$\mathbb E_G[v_{(1)}]=\mu$.
Let $a^*\in(0,1)$ be the unique solution to
\[
a^*(1-\ln a^*)=\mu.
\]
Consider the joint distribution $G^*$ under which
bidder 1 follows the equal-revenue distribution
\[
F_{a^*}(x)
=
\begin{cases}
0, & x<a^*,\\[2pt]
1-\dfrac{a^*}{x}, & a^*\leq x<1,\\[4pt]
1, & x=1,
\end{cases}
\]
while all other bidders have value zero.
Since
$\mathbb E_{G^*}[v_{(1)}]
=a^*(1-\ln a^*)=\mu$,  $G^*\in\mathcal G_1(\mu)$.

We next define the candidate mechanism. Let $H^*$
be the reserve distribution given by
\[
H^*(x)
=
\begin{cases}
0, & x<a^*,\\[2pt]
1-\dfrac{\ln x}{\ln a^*},
& a^*\leq x\leq1.
\end{cases}
\]
Let $D^*$ denote the SPA-RR with reserve distribution $H^*$.

\setcounter{theorem}{3}
\begin{theorem}\label{thm:corr-highest}
When bidders' values may follow an arbitrary joint
distribution and the seller knows $\mathbb E_G[v_{(1)}]=\mu$, robust DIC--BIC
equivalence holds. The mechanism $D^*$ is robustly
optimal, and the joint distribution $G^*$ is worst-case:
\[
\begin{aligned}
\inf_{G\in\mathcal G_1(\mu)}
\operatorname{Rev}(D^*,G)
=
\underline R_{\mathrm{corr}}^{\mathrm{DIC}}(\mu)
=
\overline R_{\mathrm{corr}}^{\mathrm{BIC}}(\mu)
=
\operatorname{OPT}^{\mathrm{BIC}}(G^*).
\end{aligned}
\]
\end{theorem}

\begin{proof}[Proof of Theorem~\ref{thm:corr-highest}]
For the upper bound, consider $G^*$.  Only bidder 1
has a  positive value. The environment
therefore reduces to a single-bidder monopoly
problem with an equal-revenue distribution,
whose optimal Bayesian revenue is $a^*$.
Hence
$\overline R_{\mathrm{corr}}^{\mathrm{BIC}}(\mu)
\leq a^*$.

For the lower bound, consider $D^*$.
For every realized value profile and reserve $r$,
the revenue equals $\max\{v_{(2)}, r\}\cdot\mathbf 1\{v_{(1)}\geq r\}$ and is therefore at least
$r\cdot\mathbf 1\{v_{(1)}\geq r\}$.
Thus, for any $G\in\mathcal G_1(\mu)$,
\[
\begin{aligned}
\operatorname{Rev}(D^*,G)
&\geq
\int_{a^*}^1
x\text{Pr}_G(v_{(1)}\geq x)\,dH^*(x)\\
&=
-\frac{1}{\ln a^*}
\int_{a^*}^1
\text{Pr}_G(v_{(1)}\geq x)\,dx.
\end{aligned}
\]
Since
$\int_0^1\Pr_G(v_{(1)}\geq x)\,dx=\mu$
and $\Pr_G(v_{(1)}\geq x)\leq1$,
\[
\int_{a^*}^1
\text{Pr}_G(v_{(1)}\geq x)\,dx
\geq
\mu-a^*
=
-a^*\ln a^*.
\]
It follows that
$\operatorname{Rev}(D^*,G)\geq a^*$
for every feasible $G$.
Combining this lower bound with the Bayesian
upper bound establishes the result.
\end{proof}

In the worst case, Nature eliminates
bidder competition.  The same revenue guarantee can also be attained
by a randomized posted-price mechanism that draws
$r\sim H^*$, allocates the good to one of the
bidders willing to pay $r$, if any, and charges
the winner $r$.\footnote{The random price distribution coincides with that in \citet{carrasco2018optimal} for a single buyer when the seller knows the mean and an upper bound on valuations. }

\subsection{Knowing the Expected Second-Highest Value}

Suppose next that the seller knows 
$\mathbb E_G[v_{(2)}]=\eta$. Recall the equal-revenue
distribution $F_{a_\eta}$ from Section~\ref{sec:known-second}.
Consider the
i.i.d.\ joint distribution $G_\eta^*$ with marginal
distribution $F_{a_\eta}$. Since
$\mathbb E_{G_\eta^*}[v_{(2)}]
=1-(1-a_\eta)^n=\eta$,
we have $G_\eta^*\in\mathcal G_2(\eta)$.

\begin{theorem}\label{thm:corr-second}
When bidders' values may follow an arbitrary joint
distribution and the seller knows 
$\mathbb E_G[v_{(2)}]=\eta$, robust DIC--BIC
equivalence holds. The standard SPA is robustly
optimal, and the joint distribution $G_\eta^*$ is worst-case:
\[
\begin{aligned}
\inf_{G\in\mathcal G_2(\eta)}
\operatorname{Rev}(\mathrm{SPA},G)
=
\underline R_{\mathrm{corr}}^{\mathrm{DIC}}(\eta)=
\overline R_{\mathrm{corr}}^{\mathrm{BIC}}(\eta)
=
\operatorname{OPT}^{\mathrm{BIC}}(G_\eta^*).
\end{aligned}
\]
\end{theorem}

\begin{proof}[Proof of Theorem~\ref{thm:corr-second}]
For every $G\in\mathcal G_2(\eta)$, the standard SPA
earns
$\operatorname{Rev}(\mathrm{SPA},G)
=\mathbb E_G[v_{(2)}]=\eta$.
Hence
$\underline R_{\mathrm{corr}}^{\mathrm{DIC}}(\eta)
\geq\eta$. For the matching upper bound, the equal-revenue
property of $F_{a_\eta}$ implies
\[
\operatorname{OPT}^{\mathrm{BIC}}(G_\eta^*)
=
1-(1-a_\eta)^n
=
\eta.
\]
Since $G_\eta^*$ is feasible,
$\overline R_{\mathrm{corr}}^{\mathrm{BIC}}(\eta)
\leq\eta$. Combining the two bounds proves
the result.
\end{proof}
Unlike in the expected-highest-value baseline, fixing \(\mathbb{E}[v_{(2)}]=\eta\) prevents Nature from eliminating competition, as the standard SPA guarantees revenue \(\eta\) under any feasible joint distribution. Nature can nevertheless limit the seller's ability to extract additional revenue by choosing an i.i.d. equal-revenue distribution, which leaves no gains from more sophisticated screening. The result for lower-ranked order statistics in
Corollary~\ref{cor:lower-order} also extends to arbitrary
joint distributions. The standard SPA guarantees at least
$\eta_K=\mathbb E_G[v_{(K)}]$, while the degenerate distribution
with all values equal to $\eta_K$ provides the matching
Bayesian upper bound.

\medskip
\noindent\textit{Known versus unknown correlation.}
Although \citet{cremer1988full} also establish full surplus extraction under dominant-strategy incentive compatibility, their formulation imposes interim rather than ex-post individual rationality. Thus, their result does not establish fixed-distribution DIC--BIC equivalence under our notion of DIC. Under distributional uncertainty, however, the worst-case distributions $G^*$ and $G_\eta^*$ identified above admit dominant-strategy incentive compatible, ex-post individually rational mechanisms that attain the optimal Bayesian revenue.  Consequently, the advantage of Bayesian mechanisms under known correlation disappears in the worst case.

\medskip
\noindent\textit{Asymmetric independent private values.}
Both  results also hold when bidders' values are independent but need not be identically distributed. The SPA-RR and the standard SPA in
Theorems~\ref{thm:corr-highest} and~\ref{thm:corr-second} attain their revenue guarantee at the independent distributions $G^*$ and $G_\eta^*$, respectively, which
also provide the matching Bayesian upper bounds. 
\newpage
\appendix
\section{Omitted Proofs}\label{appendix}

\subsection{Proof of Validity of \texorpdfstring{$H_a$}{H\_a}}\label{appendixv}
\noindent\textit{Boundary values.}
By construction, $H_a(x)=0$ for $x\leq a$, so
$H_a(0)=0$. Moreover, $H_a(1)=1$ by the definition
of $\lambda_a$.

\medskip
\noindent\textit{Continuity.}
Continuity on $[0,a)$ and $(a,1]$ is immediate. For $x>a$,
let $s=(x-a)/x$. The change of variables
$u=(t-a)/t$ gives
\[
\frac{H_a(x)}{\lambda_a}
=
s^{-(n-1)}
\int_0^s\frac{u^{n-1}}{1-u}\,du.
\]
Since $1/(1-u)\leq1/(1-s)$ for $u\in[0,s]$,
\[
0\leq\frac{H_a(x)}{\lambda_a}
\leq\frac{s}{n(1-s)}
\longrightarrow0
\qquad\text{as }x\downarrow a.
\]
Thus $H_a$ is continuous at $a$ and hence
on $[0,1]$.

\medskip
\noindent\textit{Monotonicity.}
Differentiating $H_a$ gives
\begin{equation}
H_a'(x)
=
\frac{\lambda_a}{x}
-
\frac{(n-1)a}{x(x-a)}H_a(x),
\qquad x\in(a,1).
\label{eq:Hprime_orderstat}\tag{A.1.1}
\end{equation}
Using the expression above and
$a/(x-a)=(1-s)/s$, we obtain
\[
\begin{aligned}
\frac{(n-1)a}{x-a}\frac{H_a(x)}{\lambda_a}
=
\frac{(n-1)(1-s)}{s^n}
\int_0^s\frac{u^{n-1}}{1-u}\,du\leq\frac{n-1}{n}<1.
\end{aligned}
\]
Hence $H_a'(x)>0$ for $x\in(a,1)$.
Together with continuity and the boundary values,
this establishes that $H_a$ is a valid CDF.

\subsection{Proof of Proposition~\ref{prop:spa-guarantee}}
\label{app:spa-guarantee}

Let $A(y):=1-y^n$ and
$B(y):=1-y^n-ny^{n-1}(1-y)$. The problem is
\[
\min_F \int_0^1 B(F(x))\,dx
\qquad\text{subject to}\qquad
\int_0^1 A(F(x))\,dx=\mu.
\]
Introducing a multiplier $\lambda$ for the informational constraint leads to the pointwise
objective $\ell(y):=B(y)-\lambda A(y)$.
We choose $\lambda$ so that $c_\mu$ satisfies the
pointwise first-order condition. Setting
$\ell'(c_\mu)=0$ yields
$\lambda_\mu=(n-1)(1-c_\mu)/c_\mu$. With this choice,
\[
\ell'(y)
=
ny^{n-2}
\left[(\lambda_\mu+n-1)y-(n-1)\right],
\]
which is negative for $y\in(0,c_\mu)$ and positive
for $y\in(c_\mu,1)$. Thus $c_\mu$ is the unique
global minimizer of $\ell$.

Since the Bernoulli distribution with probability
$c_\mu$ at zero and $1-c_\mu$ at one is feasible,
pointwise minimization implies that, for every
$F\in\mathcal F_1(\mu)$,
\[
\mathbb E_F[v_{(2)}]
\geq
B(c_\mu)
=
\underline\eta(\mu).
\]
Equality requires $F(x)=c_\mu$ almost everywhere
on $(0,1)$. By monotonicity and right-continuity
of CDFs, this uniquely determines the Bernoulli
distribution.

\subsection{Proof of Proposition \ref{prop}}\label{app}
Let $s:=\sup\operatorname{supp}(H)>0$. If $s>\eta$, consider the degenerate
value distribution at $\eta$. It belongs to $\mathcal F_2(\eta)$, while
the SPA-RR earns only $\eta H(\eta)<\eta$, because no sale occurs
whenever the realized reserve exceeds $\eta$.

Suppose now that $s\leq\eta$. For $l<s$, let $F_l$ be supported on
$\{l,1\}$, with probability $q_l$ on $1$, where $q_l\in(0,1)$ is chosen so
that $\mathbb E_{F_l}[v_{(2)}]
=
l+(1-l)\left[1-(1-q_l)^n-nq_l(1-q_l)^{n-1}\right]=\eta$. Such a $q_l$ exists uniquely, since the expression
in brackets is continuous and strictly increasing
in $q_l$ from $0$ to $1$.
Thus, $F_l\in\mathcal F_2(\eta)$. Let $P_0=(1-q_l)^n$ and $P_1=nq_l(1-q_l)^{n-1}$ denote the probabilities that zero and one bidder have value
$1$, respectively. Relative to the standard SPA, the random reserve loses $l$ in the
all-low state whenever the reserve exceeds $l$, while in the state with
exactly one high bidder it gains only the amount by which the reserve exceeds
$l$. Hence
\[
\operatorname{Rev}(H,F_l)-\eta
=
-lP_0(1-H(l))
+
P_1\int_{(l,s]}(r-l)\,dH(r).
\]
Since $\int_{(l,s]}(r-l)\,dH(r)
\leq
(s-l)(1-H(l))$, we have
\[
\operatorname{Rev}(H,F_l)-\eta
\leq
(1-H(l))
\left[-lP_0+(s-l)P_1\right].
\]
As $l\uparrow s$, $q_l$ converges to some $q_s<1$, so the expression in
brackets converges to $-s(1-q_s)^n<0$. Moreover, $1-H(l)>0$ for every
$l<s$. Therefore, for $l$ sufficiently close to $s$,
\[
\operatorname{Rev}(H,F_l)<\eta.
\]
Thus, every nontrivial use of a positive reserve strictly lowers the
revenue guarantee.
\subsection{Proof of Existence and Uniqueness of the Parameters for \texorpdfstring{$F_{a,b}$}{F\_{a,b}}}\label{app:low-parameters1}

\begin{claim}\label{clm:high-parameters}
Fix $\mu\in(0,1)$. For every
$\bar\eta(\mu)<\eta<\mu$, there exists a unique pair
$(a,b)$ with $0<a<b<1$ such that
$\mathbb E_{F_{a,b}}[v_{(1)}]=\mu$ and
$\mathbb E_{F_{a,b}}[v_{(2)}]=\eta$.
\end{claim}

\begin{proof}[Proof of Claim~\ref{clm:high-parameters}]
For $F_{a,b}$, the two informational constraints are
\begin{align}
\mu
&=
a+\int_a^b
\left[1-\left(1-\frac{a}{x}\right)^n\right]dx,
\label{eq:mu-high-competition}\tag{A.4.1}
\\
\eta
&=
b\left[1-\left(1-\frac{a}{b}\right)^n\right].
\label{eq:eta-high-competition}\tag{A.4.2}
\end{align}
Set $a=tb$. Then $\mu=bM_n(t)$ and
$\eta=bS_n(t)$, where
\[
M_n(t):=
t+\int_t^1
\left[1-\left(1-\frac{t}{u}\right)^n\right]du,
\qquad
S_n(t):=1-(1-t)^n.
\]
Thus $b=\mu/M_n(t)$ and
$\Phi_n(t):=S_n(t)/M_n(t)=\eta/\mu$.

We first show that $\Phi_n$ is strictly increasing.
Since
\[
M_n'(t)
=
n\int_t^1
\left(1-\frac{t}{u}\right)^{n-1}
\frac{du}{u}>0,
\]
$M_n$ is strictly increasing from $0$ to $1$.
In particular, $M_n(a_\mu)=\mu$.
Moreover, letting $s=1-t$ gives
\[
\frac{M_n'(t)}{S_n'(t)}
=
\frac{1}{s^{n-1}}
\int_0^s\frac{u^{n-1}}{1-u}\,du
=
s\int_0^1\frac{u^{n-1}}{1-su}\,du.
\]
The last expression is strictly increasing in $s$
and hence strictly decreasing in $t$.
Since $M_n(0)=S_n(0)=0$, the monotone
L'H\^opital rule implies that $M_n/S_n$ is
strictly decreasing, so $\Phi_n$ is strictly
increasing.

The restriction $b<1$ is equivalent to
$t> a_\mu$. Since
$\Phi_n(a_\mu)=\bar\eta(\mu)/\mu$ and
$\Phi_n(1)=1$, every
$\bar\eta(\mu)<\eta<\mu$ uniquely determines
$t\in(a_\mu,1)$ satisfying
$\Phi_n(t)=\eta/\mu$.
Setting $b=\mu/M_n(t)$ and $a=tb$ then gives
$0<a<b<1$ and satisfies both informational
constraints. Uniqueness follows from the strict
monotonicity of $\Phi_n$.
\end{proof}

\subsection{Proof of Existence and Uniqueness of the Parameters for \texorpdfstring{$F_{a,c}$}{F\_{a,c}}}\label{app:low-parameters2}
\begin{claim}\label{clm:low-parameters}
Fix $\mu\in(0,1)$. For every
$\underline\eta(\mu)<\eta<\bar\eta(\mu)$,
there exists a unique pair $(a,c)$ with
$0<a<1$ and $0<c<1$ such that
$\mathbb E_{F_{a,c}}[v_{(1)}]=\mu$ and
$\mathbb E_{F_{a,c}}[v_{(2)}]=\eta$.
\end{claim}

\begin{proof}[Proof of Claim~\ref{clm:low-parameters}]
For $F_{a,c}$, the two informational constraints are
\begin{align}
\mu
&=
1-(1-k)^n
+
nk\int_a^1
\left(\frac{x-k}{x}\right)^{n-1}
\frac{dx}{x},
\label{eq:low-mu}\tag{A.5.1}
\\
\eta
&=
1-(1-k)^n-nkc^{n-1},
\label{eq:low-eta}\tag{A.5.2}
\end{align}
where $k=(1-c)a$.
Set $a=k/(1-c)$ and define
$R(k):=1-(1-k)^n$ and
\[
I(k,c):=
\int_c^{1-k}\frac{u^{n-1}}{1-u}\,du.
\]
The constraints become
$\mu=R(k)+nkI(k,c)$ and
$\eta=R(k)-nkc^{n-1}$.

For fixed $c$, the first constraint is strictly
increasing in $k$, since
$\partial\mu/\partial k=nI(k,c)>0$
for $0<k<1-c$. As $k$ increases from $0$
to $1-c$, the first constraint increases
continuously from $0$
to $1-c^n$. Recall that $c_\mu=(1-\mu)^{1/n}$.
Thus, for each $c\in[0,c_\mu]$, there is a
unique $k=k_\mu(c)\in(0,1-c]$ satisfying
the first constraint, and no such $k$ exists
for $c>c_\mu$.

Along this curve, define
$\Psi_n(c):=R(k_\mu(c))-nk_\mu(c)c^{n-1}$.
We show that $\Psi_n$ is strictly decreasing.
Implicit differentiation of the first constraint
gives
\[
k_\mu'(c)
=
\frac{kc^{n-1}}{(1-c)I(k,c)},
\]
where $k=k_\mu(c)$. Consequently,
\[
\Psi_n'(c)
=
nk c^{n-2}
\left[
\frac{c[(1-k)^{n-1}-c^{n-1}]}
{(1-c)I(k,c)}
-(n-1)
\right]<0,
\]
where the inequality follows from
\[
\begin{aligned}
&(n-1)(1-c)I(k,c)
-c[(1-k)^{n-1}-c^{n-1}]\\
&\qquad =
(n-1)\int_c^{1-k}
\frac{u^{n-2}(u-c)}{1-u}\,du>0
\end{aligned}
\]
for $c\in(0,c_\mu)$.

At $c=0$, the distribution reduces to $F_{a_\mu}$,
so $k_\mu(0)=a_\mu$ and
$\Psi_n(0)=\bar\eta(\mu)$.
At $c=c_\mu$, we have $k_\mu(c_\mu)=1-c_\mu$,
giving
$\Psi_n(c_\mu)=\underline\eta(\mu)$.
Thus $\Psi_n$ maps $[0,c_\mu]$ continuously
and strictly decreasingly onto
$[\underline\eta(\mu),\bar\eta(\mu)]$.

For every
$\underline\eta(\mu)<\eta<\bar\eta(\mu)$,
there is therefore a unique $c\in(0,c_\mu)$
such that $\Psi_n(c)=\eta$. The first constraint
uniquely determines $k=k_\mu(c)$, and hence
$a=k/(1-c)$. Since $c<c_\mu$, the solution satisfies
$0<k<1-c$, implying $0<a<1$. This establishes
existence and uniqueness.
\end{proof}
\subsection{Proof of Validity of \texorpdfstring{$H_{a,c}$}{H\_{a,c}}}\label{appv2}
\noindent\textit{Boundary values.}
By construction, $H_{a,c}(0)=0$, while
$H_{a,c}(1)=1$ by the definition of $\lambda_{a,c}$.

\medskip
\noindent\textit{Continuity.}
Continuity on $[0,a)$ and $(a,1]$ is immediate. Since
$a-k=ac>0$, the prefactor and integrand in
the definition of $H_{a,c}$ remain bounded
as $x\downarrow a$, while the integral tends
to zero. Hence
$\lim_{x\downarrow a}H_{a,c}(x)=0=H_{a,c}(a)$,
so $H_{a,c}$ is continuous on $[0,1]$.

\medskip
\noindent\textit{Monotonicity.}
Differentiating $H_{a,c}$ gives
\begin{equation}
H_{a,c}'(x)
=
\frac{\lambda_{a,c}(x-a)-(n-1)kH_{a,c}(x)}
{x(x-k)},
\qquad x\in(a,1).
\label{eq:Hac-derivative}\tag{A.6.1}
\end{equation}
Let $Q(x):=\lambda_{a,c}(x-a)/[(n-1)k]$
and $D(x):=Q(x)-H_{a,c}(x)$.
Then $D(a)=0$, and \eqref{eq:Hac-derivative}
implies
\[
D'(x)+\frac{(n-1)k}{x(x-k)}D(x)
=
\frac{\lambda_{a,c}}{(n-1)k}>0.
\]
Multiplying by the integrating factor
$m(x):=((x-k)/x)^{n-1}$ and integrating
from $a$ to $x$ yields
\[
m(x)D(x)
=
\frac{\lambda_{a,c}}{(n-1)k}
\int_a^x m(t)\,dt>0
\qquad \text{for }x>a.
\]
Hence $D(x)>0$, and \eqref{eq:Hac-derivative}
gives $H_{a,c}'(x)>0$ on $(a,1)$.
Together with continuity and the boundary values,
this establishes that $H_{a,c}$ is a valid CDF.
\subsection{Proof of Theorem~3(ii)}
\label{app:thm3}
As shown in Section~\ref{sec:low-competition},
$\operatorname{OPT}^{\mathrm{BIC}}(F_{a,c})=R_{a,c}$,
so $\overline R^{\mathrm{BIC}}(\mu,\eta)\leq R_{a,c}$.
We now establish the matching lower bound.

\medskip
\noindent\textbf{Step 1: Set up the constrained minimization.}

By Lemma~\ref{lrp}, for an absolutely continuous
reserve distribution $H$, we have
\[
\operatorname{Rev}(H,F)
=
\int_0^1
\left[
(1-F(x)^n)\bigl(xH'(x)+H(x)\bigr)
-
nH(x)F(x)^{n-1}(1-F(x))
\right]dx.
\]
Introduce multipliers $\lambda$ and $\gamma$ for
the constraints $\mathbb E[v_{(1)}]=\mu$ and
$\mathbb E[v_{(2)}]=\eta$. The Lagrangian is
$\mathcal L(F;H,\lambda,\gamma)
=\int_0^1\ell_x(F(x))\,dx+\lambda\mu+\gamma\eta$,
where
\[
\begin{aligned}
\ell_x(y)
={}&
(1-y^n)\bigl(xH'(x)+H(x)-\lambda-\gamma\bigr)-n(H(x)-\gamma)y^{n-1}(1-y).
\end{aligned}
\]
We construct $H,\lambda,\gamma$ such that
$F_{a,c}(x)$ minimizes $\ell_x$ pointwise.

\medskip
\noindent\textbf{Step 2: Construct the multipliers.}

As in the proof of Theorem~1, we use the envelope
derivatives of the candidate revenue with respect
to the informational constraints to construct
the multipliers. The desired inequality will be
verified directly in Step~4.

Recall from Appendix~A.5 that, with $k=(1-c)a$,
the informational constraints can be expressed as
$\mu(k,c)=R(k)+nkI(k,c)$ and
$\eta(k,c)=R(k)-nkc^{n-1}$,
where $R$ and $I$ are defined there.
Using
$\partial_k I=-(1-k)^{n-1}/k$ and
$\partial_c I=-c^{n-1}/(1-c)$, differentiation gives
\[
\begin{aligned}
\partial_k\mu
&=nI(k,c),
&
\partial_c\mu
&=-\frac{nk c^{n-1}}{1-c},\\
\partial_k\eta
&=n\bigl[(1-k)^{n-1}-c^{n-1}\bigr],
&
\partial_c\eta
&=-nk(n-1)c^{n-2}.
\end{aligned}
\]

Let $\widehat R(\mu,\eta)$ denote the candidate
revenue expressed in terms of the two informational
constraints, and set
$\lambda=\partial_\mu\widehat R$ and
$\gamma=\partial_\eta\widehat R$.
Since
$R(k)=\widehat R(\mu(k,c),\eta(k,c))$,
the chain rule yields
\[
0=\lambda\cdot\partial_c\mu+\gamma\cdot\partial_c\eta,
\qquad
n(1-k)^{n-1}
=\lambda\cdot\partial_k\mu+\gamma\cdot\partial_k\eta.
\]
The first identity gives
\begin{equation}
\gamma
=
-\frac{c}{(n-1)(1-c)}\lambda.
\label{eq:low-gamma}\tag{A.7.1}
\end{equation}
Substituting into the second identity yields
\[
\lambda
=
\frac{(1-k)^{n-1}}
{I(k,c)-\dfrac{c}{(n-1)(1-c)}
\bigl[(1-k)^{n-1}-c^{n-1}\bigr]}.
\]
Using $k=a(1-c)$ and a change of variables, the denominator equals
$\int_a^1
\left(\frac{t-k}{t}\right)^{n-1}
\frac{t-a}{t-k}\frac{dt}{t}$. Hence
\begin{equation}
\lambda
=
\frac{(1-k)^{n-1}}
{\displaystyle
\int_a^1
\left(\frac{t-k}{t}\right)^{n-1}
\frac{t-a}{t-k}\frac{dt}{t}}
>0.
\label{eq:low-lambda}\tag{A.7.2}
\end{equation}
Denote the resulting multipliers by $\lambda_{a,c}$ and $\gamma_{a,c}$, respectively.

\medskip
\noindent\textbf{Step 3: Construct the reserve distribution.}

Differentiating the pointwise objective gives
\begin{equation}
\frac{\partial\ell_x}{\partial y}
=
ny^{n-2}
\left\{
\bigl[(n-1)(H(x)-\gamma_{a,c})-xH'(x)+\lambda_{a,c}\bigr]y
-
(n-1)(H(x)-\gamma_{a,c})
\right\}.
\label{eq:low-derivative}\tag{A.7.3}
\end{equation}
Set $H(x)=0$ for $x\leq a$. For $x\in(a,1)$,
the first-order condition at
$y=F_{a,c}(x)=1-k/x$, together with
\eqref{eq:low-gamma}, gives
\begin{equation}
xH'(x)+\frac{(n-1)k}{x-k}H(x)
=
\lambda_{a,c}\frac{x-a}{x-k}.
\label{eq:low-ode}\tag{A.7.4}
\end{equation}
Solving this equation with $H(a)=0$ yields
\begin{equation}
H(x)
=
\lambda_{a,c}
\left(\frac{x}{x-k}\right)^{n-1}
\int_a^x
\left(\frac{t-k}{t}\right)^{n-1}
\frac{t-a}{t-k}\frac{dt}{t},
\qquad x\in(a,1].
\label{eq:low-reserve}\tag{A.7.5}
\end{equation}
Denote it by $H_{a,c}$.

\medskip
\noindent\textbf{Step 4: Verify pointwise minimization.}

For $x<a$, since $H_{a,c}(x)=H'_{a,c}(x)=0$,
equations \eqref{eq:low-gamma}  and \eqref{eq:low-derivative} give
\[
\frac{\partial\ell_x}{\partial y}
=
\frac{n\lambda_{a,c}}{1-c}y^{n-2}(y-c).
\]
Thus $\ell_x$ is uniquely minimized at
$y=c=F_{a,c}(x)$.

For $x\in(a,1)$, using
\eqref{eq:low-ode} in \eqref{eq:low-derivative}
gives
\[
\frac{\partial\ell_x}{\partial y}
=
\frac{n(n-1)(H_{a,c}(x)-\gamma_{a,c})}
{F_{a,c}(x)}
y^{n-2}\bigl(y-F_{a,c}(x)\bigr).
\]
Since $H_{a,c}(x)\geq0$ and $\gamma_{a,c}<0$,
$\ell_x$ is again uniquely minimized at
$y=F_{a,c}(x)$. Integrating the pointwise
inequalities and using the two informational
constraints yields
\[
\operatorname{Rev}(D_{a,c},F)
\geq
\operatorname{Rev}(D_{a,c},F_{a,c})
\qquad
\text{for every }F\in\mathcal F_{1,2}(\mu,\eta).
\]

It remains to evaluate the expected revenue under $F_{a,c}$.
Since $H_{a,c}(x)=0$ for $x\leq a$ and
$F'_{a,c}(x)=(1-F_{a,c}(x))/x$ for $x\in(a,1)$,
the revenue integrand in Lemma~1 is the derivative of
$xH_{a,c}(x)(1-F_{a,c}(x)^n)$.
Therefore,
\[
\begin{aligned}
\operatorname{Rev}(D_{a,c},F_{a,c})
&=
\left[
xH_{a,c}(x)(1-F_{a,c}(x)^n)
\right]_{a}^{1^-}\\
&=1-(1-k)^n=R_{a,c},
\end{aligned}
\]
where $H_{a,c}(a)=0$, $H_{a,c}(1^-)=1$,
and $F_{a,c}(1^-)=1-k$.

Consequently,
\[
R_{a,c}
\leq
\underline R^{\mathrm{DIC}}(\mu,\eta)
\leq
\overline R^{\mathrm{BIC}}(\mu,\eta)
\leq R_{a,c}.
\]
All inequalities therefore hold with equality,
proving Theorem~3(ii).

\bibliographystyle{apalike}
\bibliography{abc}

\end{document}